\documentclass{Interspeech}
\usepackage[T1]{fontenc}
\usepackage[utf8]{inputenc}
\usepackage{amsmath,amssymb}  
\usepackage{graphicx}                  
\usepackage{caption}
\usepackage{comment}  
\usepackage{subcaption}
\usepackage{hyperref}                  
\usepackage{lineno}
\usepackage{xcolor}

\usepackage{comment}

 \interspeechcameraready

\title{A High-Fidelity Speech Super Resolution Network using a Complex Global Attention Module with Spectro-Temporal Loss}

\author[affiliation={1}]{Tarikul Islam}{Tamiti}
\author[affiliation={1}]{Biraj}{Joshi}
\author[affiliation={1}]{Rida}{Hasan}
\author[affiliation={2}]{Rashedul}{Hasan}
\author[affiliation={2}]{Taieba}{Athay}
\author[affiliation={2}]{Nursad}{Mamun}
\author[affiliation={1}]{Anomadarshi}{Barua}

\affiliation{Cyber-Security Engineering}{George Mason University}{USA}
\affiliation{Electronics and Telecommunication Engineering}{Chittagong University of Engineering \& Technology}{Bangladesh}
\email{abarua8@gmu.edu}

\keywords{Speech Super-resolution, Complex-valued Network, Complex Global Attention}

\begin{document}

\maketitle


\begin{abstract}

Speech super-resolution (SSR) enhances low-resolution speech by increasing the sampling rate. While most SSR methods focus on magnitude reconstruction, recent research highlights the importance of phase reconstruction for improved perceptual quality. Therefore, we introduce CTFT-Net, a Complex Time-Frequency Transformation Network that reconstructs both magnitude and phase in complex domains for improved SSR tasks. It incorporates a complex global attention block to model inter-phoneme and inter-frequency dependencies and a complex conformer to capture long-range and local features, improving frequency reconstruction and noise robustness. CTFT-Net employs time-domain and multi-resolution frequency-domain loss functions for better generalization. Experiments show CTFT-Net outperforms state-of-the-art models (NU-Wave, WSRGlow, NVSR, AERO) on the VCTK dataset, particularly for extreme upsampling (2 kHz to 48 kHz), reconstructing high frequencies effectively without noisy artifacts.

\end{abstract}   

\section{Introduction}

Speech super-resolution (SSR), also known as bandwidth extension (BWE) \cite{su2021bandwidth}, generates missing high frequencies from low-frequency speech contents to improve speech clarity and naturalness. Therefore, SSR is making its way into different practical applications, where speech quality enhancement \cite{chennoukh2001speech} and text-to-speech synthesis \cite{nakamura2014mel} are required.

Recently, deep neural networks (DNNs) became the state-of-the-art (SOTA) solutions for SSR, that operate on raw waveforms in time domains \cite{kim2024audio, li2021real, su2021bandwidth, han2022nu} or in full spectral domains \cite{lagrange2020bandwidth, li2018speech, kumar2020nu, eskimez2019speech,li2015deep}. Both domains have certain advantages and disadvantages. Time-domain methods don't need phase prediction but cannot leverage the known auditory patterns from a time-frequency (T-F) spectrogram. Moreover, the length of raw waveforms, especially at high-resolution (HR), is extremely long, hence its modeling is computationally expensive in time-domains. In contrast, spectral methods cannot predict phase and hence, need a vocoder to generate audio from real-valued spectrograms. To solve the problems that exist in both domains, we propose the Complex Time-Frequency Transformation Network (CTFT-Net), which receives complex-valued T-F spectrograms at its input and generates complex-valued T-F spectrograms, subsequently converted to raw waveform at its output. Moreover, motivated by the fact that phase plays a crucial role in speech enhancement \cite{yin2020phasen}, CTFT-Net adopts joint reconstruction of frequencies and phases from complex T-F spectrograms, providing better results for SSR tasks.


We show that our proposed CTFT-Net, a U-Net style model, provides BWE from the lowest 2 kHz input resolution to 48 kHz target resolution (i.e., upsampling ratio 24) by outperforming SOTA models \cite{lee2021nu, zhang2021wsrglow, liu2022neural, mandel2023aero, yang2024sdnet} in terms of log spectral distance (LSD) without causing artifacts at the verge between existing and generated frequency bands. Moreover, the proposed model's ability to joint estimation of complex phases and frequencies resolves the following three common issues: our model  (i) does not need to utilize the unprocessed phase from the input speech \cite{li2015deep} while reconstructing speech in time-domain, (ii) does not need to reuse the low-frequency bands of the input via concatenation \cite{liu2022neural, liuaudiosr} at post-processing, and (iii) does not need to flip the existing low-resolution (LR) phase \cite{eskimez2019speech} to reconstruct in time-domain.

This paper designs a dual-path attention block in the full complex domain to capture long-range correlations along both the time and frequency axes, referred to as the complex global attention block (CGAB). The CGAB parallelly pays attention to inter-phoneme and inter-frequency dependencies in both time and frequency axes of a complex-valued spectrogram to effectively reconstruct the missing high frequencies and phases. Therefore, CTFT-Net can be termed as a cross-domain framework, which directly uses time, phase, and frequency domain metrics to supervise the network learning. Moreover,  a complex-valued conformer is integrated into the bottleneck layer of our CTFT-Net to enhance its capability to provide local and global attention among consecutive spectrograms.

\begin{figure*}[htbp]
  \centering
\includegraphics[width=0.89\textwidth,height=0.21\textheight]{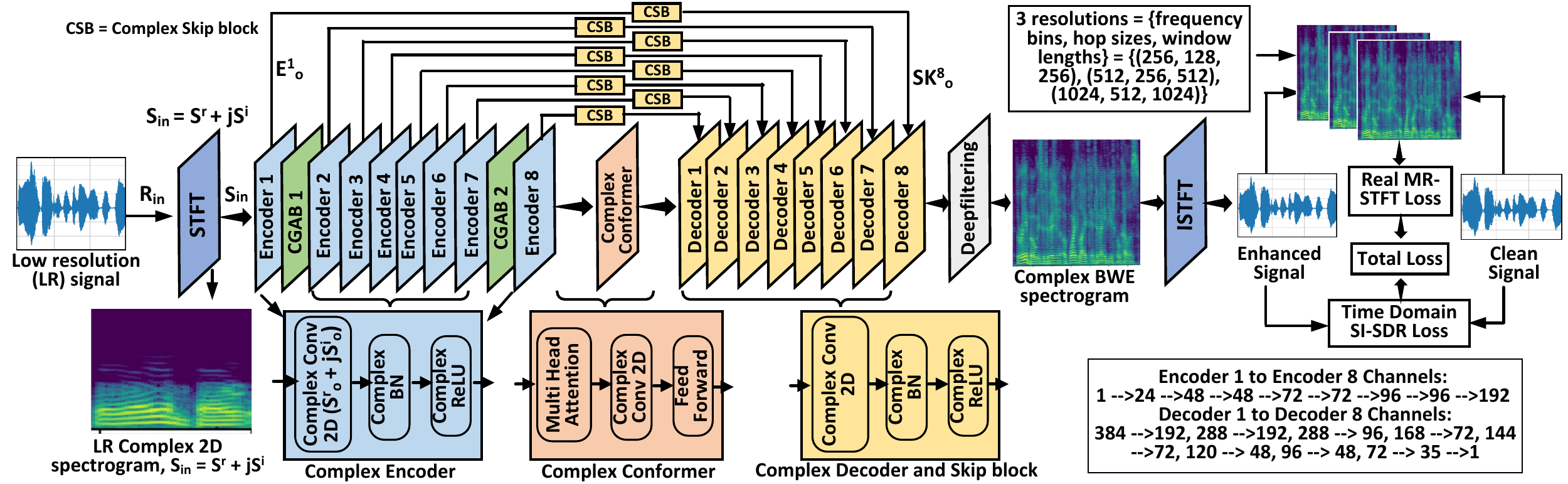}
\vspace{-0.650em}
  \caption{CTFT-Net has complex encoders, decoders, complex skip blocks, CGAB, and real MR-STFT + SI-SDR loss. }
  \label{fig:overall_architecture}
  \vspace{-01.750em}
\end{figure*}

We combine the scale-invariant signal-to-distortion ratio (SI-SDR) \cite{le2019sdr} loss with real-valued multiresolution short-time Fourier transform (STFT) loss \cite{tian2020tfgan} for joint optimization in both time and frequency domains. We show that this combination provides better results for SSR in complex domains. Experimental results show that CTFT-Net outperforms the SOTA baselines, such as NU-Wave, WSRGlow, NVSR, and AERO on LSD for the VCTK multispeaker dataset. Notably, CTFT-Net~\footnotemark performs better for extremely LR speech signals, such as upsampling from a minimum of 2 kHz to 48 kHz. 

\footnotetext{Source code of the model will be available after acceptance.}





In a nutshell, the technical contributions of our work are:
\begin{itemize} 

\item We propose a cross-domain SSR framework that operates entirely in complex domains, jointly reconstructing both magnitude and phase from the LR speech signal.

\item We propose CGAB - a dual-path end-to-end attention block in encoders and use conformers in bottleneck layers in the full complex domain to capture the long-range correlations along both the time and frequency axes. 

\item  We integrate SI-SDR loss in time-domain with multi-resolution STFT loss in the frequency domain to capture fine-grained and coarse-grained T-F spectral details.

\item We perform a comprehensive ablation study and evaluate the proposed model on the VCTK multispeaker dataset. Results show that CTFT-Net outperforms the SOTA SSR models.





\end{itemize}

\vspace{-0.5em}
\section{Methodology} 
\vspace{-0.1em}

Here, we discuss our proposed modifications on U-Net that construct complex-valued CTFT-Net for SSR tasks.

\vspace{-0.5em}
\subsection{Proposed network architecture in complex-domain}
\vspace{-0.2em}
\label{subsec:General network architecture}

The detailed architecture of the proposed CTFT-Net is shown in Fig. \ref{fig:overall_architecture}. The network consists of four main components: (i) a total of 16 (i.e., 8 + 8) full complex-valued encoder-decoder blocks, (ii) complex-valued skip blocks, (iii) complex-valued conformer in the bottleneck layer, and (iv) complex-valued global attention blocks - CGAB. The complex domain processing by our proposed CTFT-Net has the potential to adopt best practices from two different domains that are explained below.  

\textbf{Reasoning behind complex-domain models:} Existing SSR methods can be classified broadly into two domains: (i) spectral methods where real-valued T-F spectrograms are provided at model's input, and (ii) time-domain methods where raw waveforms are provided at model's input. Spectral methods cannot predict phase and hence, need a vocoder to generate audio from the bandwidth-extended spectrograms. Moreover, spectral methods typically use mean square error (MSE) loss, and cannot directly use time-domain loss functions, such as SI-SDR loss to improve the speech quality while performing BWE. In contrast, time-domain methods avoid phase prediction problems and can include SDR-type loss function but cannot leverage the known auditory patterns from a T-F spectrogram.

Our proposed CTFT-Net handles both frequencies and phases simultaneously by receiving complex T-F spectrograms at its input and generates raw waveform at its output without any vocoders. Therefore, it is typically free from the problems that both spectral and time domain methods have and can deliver superior SSR compared to the SOTA models.

\vspace{-0.5em}
\subsection{Complex encoders and decoders}
\vspace{-0.5em}
\label{subsec:complex_encoder}

Each encoder/decoder block is built upon complex-valued convolution to ensure successive extraction and reconstruction of both magnitude and phase from the complex T-F spectrogram. Complex convolution is the key difference between a complex-valued network and a real-valued network. Formally, the input LR waveform $R_{in}$ is first transformed into STFT spectrogram, denoted by $S_{in}$ in Fig. \ref{fig:overall_architecture}. Here, $S_{in} (= S^r+ jS^i) \in \mathbb{C}^{F \times T}$ is a complex-valued spectrogram, where $F$ denotes the number of frequency bins and $T$ denotes the number of time frames. $S_{in}$ is fed into 2D complex convolution layers of encoders to produce feature $S_0 \in \mathbb{C}^{F \times T \times C}$, where C is the number of channels. If complex kernel is denoted by $W = {W}_r + j{W}_i$, the complex convolution is defined as:

\vspace{-0.5em}
\begin{equation}
\begin{aligned}
{S}^r_0 &= {W}_r * {S}^r_{in} - {W}_i * {S}^i_{in} + {b}_r, \\
{S}^i_0 &= {W}_r * {S}^i_{in} - {W}_i * {S}^r_{in} + {b}_i,
\end{aligned}
\label{eq:complex_conv}
\vspace{-0.0em}
\end{equation}



where $*$ denotes the convolution, $S^r_0$ $\&$ $S^i_0$ are real and imaginary parts of  $S_0$, and ${b}_r$ $\&$ ${b}_i$ are bias terms. The convolution output is then normalized using complex batch normalization (BN) for stable training and passed through a complex ReLU activation for adding non-linearity. Formally, encoder outputs, denoted by $E^n_0$ =  $Cplx ReLU ( Cplx BN (S^r_0 + j S^i_0))$, where n = 1 to 8 and $Cplx$ refers to complex operations. Complex decoders are similar to complex encoders except complex convolution is substituted by complex-transpose convolution.

\vspace{-0.5em}
\subsection{Complex skip block} 
\label{subsec:complexskipblock}
\vspace{-0.3em}

A skip connection in our proposed CTFT-Net passes high-dimensional features from the complex-valued encoders to the appropriate decoders. This enables the model to preserve the spatial features, which may lost during the down-sampling operation, and guides the network to propagate from encoders to decoders. CTFT-Net implements skip blocks in complex domains, inspired by \cite{kothapally2020skipconvnet}, to enable the proper flow of complex features from the encoder's output to decoders. Each complex skip block applies a complex convolution on the encoder output $E^n_0$,  followed by a complex BN and a complex ReLU activation. Formally, the complex skip block's output, denoted by $SK^n_0$ =  $Cplx ReLU ( Cplx BN (Cplx Conv (E^n_0)))$, where the $CplxConv$ is implemented following Eqn. ~\eqref{eq:complex_conv}.

\vspace{-0.5em}
\subsection{Complex global attention block (CGAB)}
\label{subsec:CGAB}
\vspace{-0.3em}

Long-range correlations exist along both the time and the frequency axes in a complex T-F spectrogram. As audio is a time series signal, inter-phoneme correlations exist along the time axis. Moreover,  harmonic correlations also exist among pitch and formants along the frequency axis. As convolution kernel is limited by their receptive fields, standard convolutions cannot capture global correlations that exist in time and frequency axes in a complex T-F spectrogram. Please note that frequency transformation blocks (FTBs) \cite{yin2020phasen} don't work along both the T-F axes. Moreover, similar to dual attention blocks (DABs) \cite{tang2021joint}, T-F attention blocks are proposed for speech enhancement \cite{zhang2022time, dang2022dpt, mamun2024speech} and dereverberation tasks \cite{kothapally2022complex}. However, attention along both the T-F axes in \textit{complex T-F spectrograms} is not well explored for the SSR task, to the best of our knowledge.

\begin{figure}[ht!]
  \centering
 \includegraphics[width=0.43\textwidth,height=0.26\textheight]{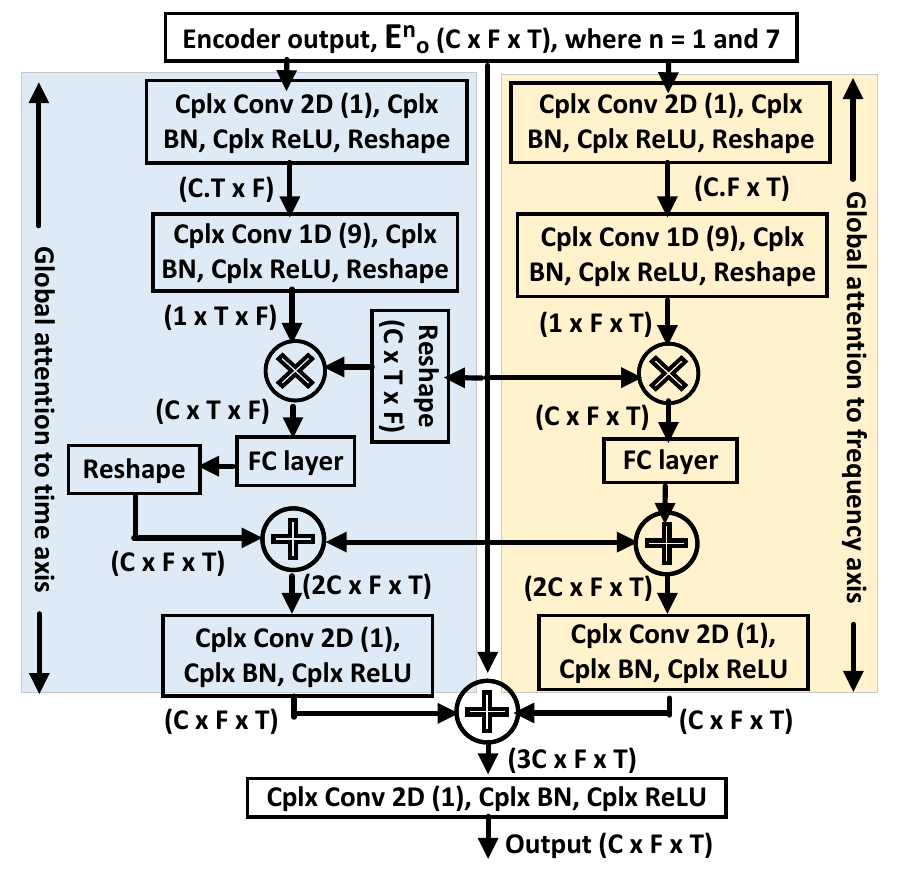} 
 \vspace{-0.905em}
  \caption{CGAB captures complex  global T-F correlations.}
  \label{fig:CGAB}
  \vspace{-01.15em}
\end{figure}

The detailed implementation of our proposed CGAB is shown in Fig. \ref{fig:CGAB}. CGAB provides attention to the time and frequency axes of a complex spectrogram by following two steps:

\textbf{Step 1 - Reshaping along the T-F axes:} The output $E^n_0$ from the encoder is decomposed in 2 steps by CGAB into two tensors: one along the time axis and another along the frequency axis. Formally, $E^n_0$, which has a feature dimension of $C \times F \times T$, is given at the input of CGAB. At the first stage of reshaping,  $E^n_0$ parallelly reshaped into $C.T$ vectors with dimension $C \cdot T \times F$ and into $C.F$ vectors with dimension $C \cdot F \times T$. This reshaping is done using 2D complex convolution, complex BN, and ReLU activation followed by vector reshaping. In the second stage of reshaping, $C \cdot T \times F$ is reshaped into $1 \times T \times F$ and  $C \cdot F \times T$ is reshaped into $1 \times F \times T$ using 1D complex convolution, complex BN, ReLU activation followed by vector reshaping. The tensors with dimension $1 \times F \times T$ capture the global harmonic correlation along the frequency axis and $1 \times T \times F$ capture the global inter-phoneme correlation along the time axis. The captured features along the T-F axes and the original features from $E^n_0$ are point-wise multiplied together to generate a combined feature map with a dimension of $C \times T \times F$ and $C \times F \times T$ along T and F axes, respectively. This point-wise multiplication captures the inter-channel relationship between the encoder's output $E^n_0$ and complex time and frequency axes.

\textbf{Step 2 - Global attention along the T-F axes:} It is possible to treat the spectrogram as a 2D image and learn the correlations between every two pixels in the 2D image. However, this is computationally too costly and is not realistic. On the other hand, ideally, we can use self-attention \cite{ashish2017attention} to learn the attention map from two consecutive complex T-F spectrograms. But this might not be necessary. Because, on the time axis in each T-F spectrogram, when calculating signal-to-noise ratio (SNR), the same set of parameters in recursive relation are used, which suggests that temporal correlation is time-invariant among consecutive spectrograms. Moreover, harmonic correlations are independent in the consecutive spectrograms \cite{scalart1996speech}.

Based on this understanding, we propose a self-attention technique along the T-F axes within each spectrogram, without considering correlations among consecutive spectrograms at this stage (see Section \ref{subsec:conformer_symbolic}). Specifically, attention on frequency and time axes are implemented by two separate fully connected (FC) layers. Along the time path, the input and output dimensions of FC layers are $C \times T \times F$. Along the frequency path, the input and output dimensions of FC layers are $C \times F \times T$. FC layer learns weights from complex T-F spectrograms and technically is different from the self-attention \cite{ashish2017attention} operation. To capture interchannel relationships among the input $E^n_0$ and output of FC layers, concatenation happens followed by 2D complex convolutions, complex BN, and complex ReLU activation. Finally, the learned weights from the T-F axes are concatenated together to form a unified tensor, which holds joint information on the T-F global correlations from each spectrogram. 

We use only two CGABs - one in between the 1st and 2nd encoders, and another one in between the 7th and 8th encoders. 

\vspace{-0.5em}
\subsection{Complex conformer in the bottleneck layer}
\label{subsec:conformer_symbolic}
\vspace{-0.3em}

We use complex-valued conformers in the bottleneck layer of our CTFT-Net to capture both local and global dependencies \textit{among consecutive spectrograms}. Our complex conformer comprises complex multi-head self-attention, complex feed-forward, and complex convolutional modules, inspired by \cite{gulati2020conformer}. The complex conformer optimally balances global context with fine-grained local information for BWE. 

\vspace{-0.5em}
\subsection{ Real Multiresolution STFT Loss + SI-SDR loss}
\label{subsec:Complex Multi-Resolution STFT Loss}
\vspace{-0.3em}

Unlike mean square error (MSE) loss \cite{liu2022neural}, we propose multiresolution STFT (MR-STFT) loss only on the real part of STFT over multiple resolutions. At first, spectral convergence loss $L_{SC}$ \cite{tian2020tfgan} and log STFT magnitude loss $L_{mag}$ \cite{tian2020tfgan} are calculated on both the real and imaginary parts of the input signal's STFT data. Let's define the $L_{SC}$ and $L_{mag}$ calculated on  real and imaginary STFT data as \{$L^r_{SC}$, $L^i_{SC}$\} and \{$L^r_{mag}$, $L^i_{mag}$\}, respectively. Assuming we have $S$  different STFT resolutions, we aggregate only the $L^r_{SC}$ and $L^r_{mag}$ over $S$ resolutions. We define this as real MR-STFT loss, $L^r_{\mathrm{MR-STFT}}$,  which is: 

\vspace{-01.8em}
{
\begin{align}
\vspace{-0.5em}
  L^r_{\mathrm{MR-STFT}} = \frac{1}{S} \sum_{s=1}^{S} \Big( L_{\mathrm{SC}}^{r} + L_{\mathrm{mag}}^{r} \Big);
\end{align}
}
\vspace{-0.85em}

We use $S$ = 3 different resolutions, such as \{frequency bins, hop sizes, window lengths\} = \{(256, 128, 256), (512, 256, 512), (1024, 512, 1024)\} to calculate $L^r_{MR-STFT}$. 
As CTFT-Net can directly generate raw waveform at its output from complex T-F spectrograms, we add SI-SDR loss \cite{le2019sdr}, $L_{SISDR}$, with  $L^{r}_{MR-STFT}$ to calculate total loss (i.e., $L^{r}_{MR-STFT} + L_{SISDR}$), improving the audio quality in both T-F domains. This joint optimization in the complex T-F domain improves the \textit{perceptual quality} of the bandwidth-extended speech. We refer to Section \ref{subsec:ablation} to understand how different losses, such as real-valued single resolution STFT loss and MR-STFT loss influence our complex-valued model.

\vspace{-0.5em}
\section{Experiments}
\label{sec:experiments}
\vspace{-0.3em}

\subsection{Speech corpus and preprocessing} 
\label{subsec:Speech_Corpus}
\vspace{-0.3em}

We use VCTK (version 0.92) \cite{yamagishi2019cstr}, a multi-speaker English corpus containing 110 speakers, for training (i.e., 95 speakers) and testing (i.e., 11 speakers). Each audio clip has a duration ranging from 2s to 7s. We standardize all audio clips to 4s by either zero-padding or trimming. Following \cite{kumar2020nu}, only the mic1 microphone data is used for experiments, and p280 and p315 are omitted for the technical issues. For the LR simulation process, we apply a sixth-order low-pass filter to prevent aliasing and then downsample the original audio from 48 kHz to different low-sampling frequencies to generate LR samples. We also use sinc interpolation to upsample before the BWE to ensure the system input and output have the same shape. 




\vspace{-0.5em}
\subsection{Training, hardware and hyperparameter details}
\label{subsec:Training_details}
\vspace{-0.3em}

Training data pairs are built and stored for faster processing. 
Training, testing, and validation are done in PyTorch Lightning. Key training parameters include a batch size of 8 with 100 epochs, and the Adam optimizer with a learning rate of \(1\times10^{-4}\), weight decay of \(1\times10^{-5}\), and momentum parameters \(\beta_1=0.5\) and \(\beta_2=0.999\). The learning rate is scheduled using Cosine Annealing Warm Restarts (with \(T_0=10\) and \(T_{\text{mult}}=1\)), gradient clipping (max norm of 10) and gradient accumulation (over 2 batches) to ensure stability. Training is executed in 32-bit precision on GPUs, utilizing a distributed data-parallel strategy. Our experiments use AMD~Ryzen\texttrademark\ 7950X3D processor (16~cores, 32~threads), 192\,GB~of~RAM,  four NVIDIA\textregistered\ RTX~4090~GPUs, and 10 TB storage.


\vspace{-0.5em}
\subsection{Comprehensive evaluation metrics}
\label{subsec:Evaluation_Metrics}
\vspace{-0.3em}

To comprehensively evaluate the reconstructed audio, we use four evaluation metrics: log spectral distance (LSD) \cite{liu2022neural} for spectral distortion, short-time objective intelligibility (STOI) \cite{taal2011algorithm} for intelligibility, perceptual evaluation of speech quality (PESQ) \cite{rix2001perceptual} for perceived quality, and scale-invariant signal-to-distortion ratio (SI-SDR) \cite{le2019sdr} for overall signal distortion.

\vspace{-0.3em}
\section{Results}
\label{sec:results}
\vspace{-0.3em}

We conduct comprehensive evaluations of CTFT-Net by comparing it with SOTA models, followed by an ablation study. 

\vspace{-0.7em}
\subsection{Performance analysis}
\label{subsec:results}
\vspace{-0.3em}

\begin{table}[ht]
\vspace{-0.98em}
\footnotesize
    \centering
    \caption{ LSD Comparison for 48 kHz target sampling rate.}
    \vspace{-0.95em}
    \begin{tabular}{l ccccc }
        \toprule
        \textbf{Model} & \textbf{2 kHz} & \textbf{4 kHz} & \textbf{8 kHz} & \textbf{12 kHz} & \textbf{Size (M)}\\
        \midrule
        Unprocessed  & 3.06& 2.85& 2.44&   1.34 & - \\
        NU-Wave \cite{lee2021nu}    & 1.85 & 1.48 & 1.45 & 1.27 & 3 \\
        WSRGlow \cite{zhang2021wsrglow}   & 1.45 & 1.18 & 1.02 & 0.91 & - (40)\\
        NVSR \cite{liu2022neural}    & 1.10 & 0.99 & 0.93 & 0.87 & 99\\
        AERO \cite{mandel2023aero}     & 1.15 & 1.09 & 1.01 & 0.93 & - (13)\\
        AP-BWE \cite{lu2024towards} & 1.016 & 0.92 & 0.84 & 0.78& - (5)\\
        \textbf{Proposed} & \textbf{1.06} & \textbf{0.96} & \textbf{0.81} &  \textbf{0.62} & 61.6 \\
        \bottomrule
    \end{tabular}
    \label{tab:comparison with SOTA}
    \vspace{-0.5em}
\end{table}
\vspace{-0.5em}

\textbf{Comparison with baselines:} We reproduced NU-Wave, NVSR, AERO, and WSRGlow for baselines with their open-sourced code \cite{nuwave,NVSR,aero,WSRGlow} and default settings for 48 kHz target from 2, 4, 8, and 12 kHz input (see Table \ref{tab:comparison with SOTA}). For each LR input, CTFT-Net achieves the lowest LSD compared to all baselines. Please note that NVSR \cite{liu2022neural} \textit{copies the LR spectrum directly} to the output in post-processing steps. CTFT-Net outperforms the baselines without any NVSR-style post-processing. Moreover, NVSR largely relies on the neural vocoder, which may become the bottleneck of NVSR's performance.  CTFT-Net does not need any vocoder as it can handle magnitude and phase jointly. \textit{Hence, CTFT-Net is objectively improved with respect to the best-evaluated baseline – NVSR.} The improvement is significant for all the input LR frequencies.  This is an indication of CTFT-Net's strength, which basically comes because of joint attention on complex T-F domains and joint optimization using real-valued MR-STFT and SI-SDR losses.

\begin{table}[ht]
\vspace{-0.5em}
\centering
\caption{CTFT-Net evaluation for target 16 kHz sampling rate.}
\vspace{-0.7em}
 \resizebox{\columnwidth}{!}{%
\begin{tabular}{l c c c c c c}
\toprule
\textbf{Bandwidth} & \multicolumn{2}{c}{\textbf{2 to 16 kHz}} 
                & \multicolumn{2}{c}{\textbf{4 to 16 kHz}}
                & \multicolumn{2}{c}{\textbf{8 to 16 kHz}} \\
\cmidrule(lr){2-3}\cmidrule(lr){4-5}\cmidrule(lr){6-7}
 & \textbf{Unprocessed} & \textbf{Enhanced}
 & \textbf{Unprocessed} & \textbf{Enhanced}
 & \textbf{Unprocessed} & \textbf{Enhanced} \\
\midrule
LSD $\downarrow$       & 2.95 & 1.01 & 2.30 & 0.98 &  1.20 &  0.72  \\
STOI $\uparrow$       & 0.79&  0.79&  0.9&  0.89 & 0.99 &  0.99  \\
PESQ $\uparrow$   & 1.14 & 1.46 &  1.32 &  1.95 &  2.33 &  2.99 \\
SI-SDR $\uparrow$    & 11.37 &  11.38 & 16.66 &  16.69 &  22.6 &  22.63  \\
\bottomrule
\end{tabular}
 }
\label{tab:evaluation}
\vspace{-0.7em}
\end{table}

\textbf{Improving perceptual quality with BWE:} Table \ref{tab:evaluation} indicates that LSD is improved by $\sim$66\%, $\sim$57\%, and $\sim$40\% for 16 kHz target frequency when upsampling from 2, 4, and 8 kHz, respectively. The STOI remains quite the same for all upsampling frequencies, indicating speech intelligibility is not sacrificed for SSR tasks at hand. Additionally,
PESQ is also improved by $\sim$28\%, $\sim$47\%, and $\sim$28\% when upsampling from 2, 4, and 8 kHz, respectively, indicating the model's ability to improve perceptual quality. Improving the signal's perceptual quality while doing BWE is typically more important when the BWE is done from a very low sampling frequency of 2 kHz. 
Please note that SI-SDR is also slightly increased for all LR input in Table \ref{tab:evaluation}. It indicates that BWE by our model does not add noisy artifacts into the final output. 


\vspace{-0.5em}
\subsection{Ablation study}
\label{subsec:ablation}
\vspace{-0.3em}


\textbf{Study of the proposed CGAB:} To justify that attention over both T-F axes in a complex-valued spectrogram is better than attention over only the frequency axis, we compare the performance between FTBs \cite{yin2020phasen} and CGABs with our model.   From lines $P_1$ and $P_{6}$ of Table \ref{tab:ablation}, it is clear that the CGAB is better than the FTB for complex-valued spectrograms as a CGAB has attention on both T-F axes. Moreover, we evaluate CTFT-Net's performance by adding CGABs in each encoder (line $P_2$). This modification improves LSD slightly by 1.8\% (1.06 $\rightarrow$ 1.04) but with an increase of the model size by 31\% (61.6 million $\rightarrow$ 80.2 million). Therefore, we don't add CGABs in each encoder in our current design of CTFT-Net.

  \vspace{-0.6em}
\begin{table}[ht] 
    \centering
    \caption{Detailed ablation study for 2 - 48 kHz upsampling where M = million.}
    \vspace{-0.75em}
    \label{tab:comparison}
    \resizebox{\columnwidth}{!}{
    \begin{tabular}{lccccccc}
        \hline
          & \textbf{Model}   & \textbf{LSD $\downarrow$} & \textbf{STOI $\uparrow$} & \textbf{PESQ $\uparrow$} & \textbf{SI-SDR $\uparrow$} & \textbf{NISQA-MOS $\uparrow$} & \textbf{Size (M)$\downarrow$}  \\
        \hline
        \hline
        $P_0$ & Unprocessed & 3.06 & 0.79 & 1.11 & 11.27 &1.27 &--  \\
        \hline
        $P_1$ & w/ FTB \cite{yin2020phasen}    &1.32 &0.78 & 1.15 & 10.54 &1.02& 10.1\\
        $P_2$ & w/ CGAB in each encoder  & 1.04 & 0.81 & 1.11 & 11.42 &1.58& 80.2 \\
        \hline
        $P_3$ & w/ post-processing  & 1.03 & 0.82 & 1.16 & 10.27 &1.52& 61.6 \\
        $P_4$ & w/ snake activation  & 1.19 &  0.78 & 1.25 & 11.4 &1.43& 61.6 \\
        \hline
        $P_5$ & w/ SR-STFT loss  & 1.4 & 0.73 & 1.13 & 3.06 &1.22& 61.6 \\
        
        $P_7$ & w/ complex MR-STFT loss & 0.98 & 0.81& 1.11 &1.55&  11.47& 61.6\\
        $P_8$ & w/ transformer in bottleneck &1.001 & 0.80 & 1.27 &1.45& 11.19 &61.6\\
        $P_9$ & w/ lattice block in bottleneck & 1 & 0.81 & 1.14 &1.57& 11.47 &17\\
        $P_{10}$ & w/o SI-SDR & 0.88 &  0.84& 1.24 & 8.77 &1.45&61.9  \\
        
        $P_{11.1}$ & w/o CGAB (down-up) & 0.98 &  0.83& 1.15 & 11.53 &1.71&9.3  \\
        $P_{11.2}$ & w/o CGAB (filtering) & 0.98 &  0.87& 1.18 & 14.5 &1.84&9.3  \\
        
        $P_{12}$ & series CGAB & 1.09 &  0.79& 1.2 & 11.08 &1.52&61.6  \\
        $P_{13}$ & AP-BWE (2-48 KHz)& 1.016 & 0.84 & 1.5 & 7.38 & 4.01 & -(5)\\
        $P_{14}$ & AP-BWE (4-48 KHz)& 0.92 & 0.94& 2.32 & 12.4& 4.01 & -(5)\\
        $P_{15}$ & NU-Wave (2-48) & 1.9 & 0.74 & 1.08 & 4.77 & 1.64 & 3 \\
        $P_{16}$ & NU-Wave (4-48) & 1.49 & 0.87 & 1.47&11.12  & 2.35 & 3 \\
        $P_{17}$ & NU-Wave (8-48) & 1.75 & 0.97 &  2.07 & 15.42&  2.84& 3\\
        $P_{18}$ & NU-Wave (12-48) & 1.51 & 0.98 & 2.97  & 17.075& 2.97 & 3\\
        \hline
        \hline
        $P_{6.1}$ & Our CTFT-Net(filtering)  &  1.06 & 0.81 &  1.15 & 11.24 &1.56& 61.6  \\
        $P_{6.2}$ & Our CTFT-Net(down-up)  &  1.01 & 0.81 &  1.19 & 11.17 &1.56& 61.6  \\
        \hline
    \end{tabular}%
    }
    \label{tab:ablation}
\end{table}
\vspace{-0.9em}

\textbf{Post processing and snake activation:} We experiment with the NVSR-style post-processing technique (line $P_3$) discussed in \cite{liu2022neural}. Line $P_{3}$ indicates that CTFT-Net gives better results with the NVSR-style post-processing. However, we don't use any post-processing in our current design to prove that CTFT-Net works much better even without any post-processing. Moreover, our model gives better results with simpler ReLU activation compared to snake activation used in \cite{mandel2023aero} (see line $P_4$).

\textbf{Real single-resolution STFT (SR-STFT) loss:} We experiment with real-valued SR-STFT loss for different values of \{frequency bins, hop sizes, window lengths\}. Experiments find that the real-valued MR-STFT loss is always better compared to real-valued SR-STFT loss for CTFT-Net because MR-STFT loss can capture fine and coarse-grained details from different resolutions. $P_{5}$ shows the real-valued SR-STFT loss for \{frequency bins, hop sizes, window lengths\} = \{320, 80, 320\}. We define real-valued SR-STFT loss, $L^r_{\mathrm{SR-STFT}}$,  as:

\vspace{-01.0em}
{
\begin{align}
\vspace{-0.5em}
  L^r_{\mathrm{SR-STFT}} =  L_{\mathrm{SC}}^{r} + L_{\mathrm{mag}}^{r}
\end{align}
}
\vspace{-0.85em}

where $L_{\mathrm{SC}}^{r}$ and  $L_{\mathrm{mag}}^{r}$ are real-parts of the spectral convergence loss \cite{tian2020tfgan} and log magnitude loss \cite{tian2020tfgan}, respectively.



\textbf{Remarks:} As CTFT-Net is trained with fixed input resolutions, it is not tested other than the same input audio resolution. Moreover, CTFT-Net is not evaluated on other than speech datasets (i.e., music, etc.) as our goal is SSR.

\vspace{-0.5em}
\section{Conclusion}
\label{sec:conclusion}
\vspace{-0.3em}

This paper presents a novel SSR framework that operates entirely in complex domains, jointly reconstructing both magnitude and phase from the LR signal using global attention on T-F axes.  It shows strong performance across a wide range of input sampling rates ranging from 2 kHz to 48 kHz. For the VCTK multi-speaker benchmark, results show that CTFT-Net outperforms the SOTA SSR models.

\bibliographystyle{IEEEtran}
\bibliography{Manuscript}

\end{document}